\documentclass[12pt]{article}

\usepackage{cite}
\usepackage{epsfig}
\usepackage{amsmath,amsfonts}
\usepackage{url}

\newcommand{\sta}{{\tilde{t}_1}}

\newcommand{\mst}[1]{m_{\tilde{t_{#1}}}}

\newcommand{\neu}{\tilde{\chi}^0}

\newcommand{\mneu}[1]{m_{\tilde{\chi}^0_{#1}}}

\newcommand{\NPB}[3]{{ Nucl.~Phys.} B \textbf{#1}, #3 (#2)}   
\newcommand{\PLB}[3]{{ Phys.~Lett.} B \textbf{#1}, #3 (#2)}   
\newcommand{\PRD}[3]{{ Phys.~Rev.} D \textbf{#1}, #3 (#2)}

\newcommand{\HPA}[3]{{ Helv.~Phys.~Acta} \textbf{#1}, #3 (#2)}

\def\mathswitch#1{\relax\ifmmode#1\else$#1$\fi}
\def\mathswitchr#1{\relax\ifmmode{\mathrm{#1}}\else$\mathrm{#1}$\fi}

\newcommand{\SLASH}[2]{#1\hspace{-#2ex}/}

\newcommand{\ppslash}{\SLASH{\vec{p}}{1.}}

\newcommand{\Eslash}{\SLASH{E}{1.5}\,}
\newcommand{\Et}{\Eslash_{\rm T}}

\newcommand{\gesim}{\,\raisebox{-.3ex}{$_{\textstyle >}\atop^{\textstyle\sim}$}\,}

\newcommand{\mycaption}[1]{\caption{\sl #1}}

\begin{document}

\title{
\vspace{-2em}
\begin{flushright}
{\normalsize\sf ANL--HEP--PR--08--46 \\
EFI--08--22 \\
FERMILAB--PUB--08--269--T \\}
\end{flushright}
\vskip 30pt
Light Stop Searches at the LHC in Events with One Hard Photon or Jet and 
Missing Energy 
}

\author{
M. Carena$^{1}$, A. Freitas$^{2,3,4}$ and C.E.M. Wagner$^{3,4}$
\\[1em] 
\small \sl $^{1}$ Fermi National Accelerator Laboratory, P.O.~Box 500, Batavia,
IL 60510, USA \\
\small \sl $^{2}$ 
Department of Physics, 100 Allen Hall University of Pittsburgh, Pittsburgh, 
PA 15260, USA \\
\small \sl $^{3}$ HEP Division, Argonne National Laboratory, 9700 Cass Ave.,
Argonne, IL 60439, USA \\
\small \sl $^{4}$ Enrico Fermi Institute and Kavli Institute for
Cosmological Physics,\\[-1ex]  \small \sl  Department of Physics, University of
Chicago, 5640 S. Ellis
Ave.,\\[-1ex]  \small \sl  
Chicago, IL 60637, USA }

\date{}

\maketitle

\begin{abstract}

Low energy supersymmetric models provide a solution to the hierarchy problem 
and also have the necessary ingredients to solve two of the most outstanding
issues in cosmology: the origin of the baryon asymmetry and the source of dark
matter. In the MSSM, weak scale generation of the baryon asymmetry may  be
achieved in the presence of light stops, with masses lower than about 130~GeV.
Moreover, the proper dark matter density may be obtained in the stop--neutralino
co-annihilation region, where the stop--neutralino mass difference is smaller
than a few tens of GeV. Searches for scalar top quarks (stops) in pair
production processes at the Tevatron and at the Large Hadron Collider (LHC)
become very challenging in this region of parameters.  At the LHC, however,
light stops proceeding from the decay of gluino pairs may be identified,
provided the   gluino mass is smaller than about 900~GeV. In this article we
propose an alternative method for stop searches in the co-annihilation region,
based on the search for these particles in events with missing energy plus one
hard photon or jet. We show that this method is quite efficient and, when
complemented with ongoing Tevatron searches, allows to probe stop masses up to
about 160~GeV, fully probing the  region of parameters consistent with
electroweak baryogenesis in the MSSM.

\end{abstract}

\thispagestyle{empty}

\setcounter{page}{0}
\setcounter{footnote}{0}

\newpage


\section{Introduction}

Supersymmetric extensions of the Standard Model (SM)  provide a natural
explanation for dark matter. Stability of the lightest supersymmetric particle
(LSP) is ensured by a discrete symmetry, R-Parity. Due to the renormalization
group evolution of the supersymmetric particle masses, the LSP is naturally
neutral and weakly interacting, and therefore has an annihilation cross section
of the order of  the one necessary for the LSP to become a good dark matter
candidate.

Low energy supersymmetry also provides a possible solution to another
outstanding problem of the Standard Model, namely the generation
of the baryon-antibaryon asymmetry~\cite{EWBGreviews,PT}.  In the minimal
supersymmetric extension of the SM (MSSM), a solution to  this problem by weak
scale physics demands a light stop as well as non-vanishing CP-violating phases
in the chargino--neutralino sector. Such a light stop tends to push the lightest
CP-even Higgs mass to values below  the  current experimental bound
\cite{lephbound}, unless the heaviest stop mass is larger than about 10 TeV. All
other squarks and sleptons should also be heavy, in order to suppress the
electron and neutron electric dipole moment 
contributions~\cite{edm1,Csaba,Caltech}. 

The required low energy theory that leads to an explanation of the origin of
dark matter and the baryon asymmetry contains the SM particles, a light stop,
and light electroweak  gauginos and higgsinos, the latter being important for
the generation of the CP-violating currents contributing to the baryon
asymmetry, as well as a source of  dark matter and a light Higgs.  Although the
scenario does not lead to a prediction for the non-SM Higgs masses, large values
of these masses, above a few 100 GeV, provide a natural suppression of the
two-loop contribution to the electron electric dipole moments and help in
obtaining a sufficiently large value of the lightest CP-even Higgs mass. For
large CP-odd Higgs boson masses, the lightest CP-even Higgs will behave in a
standard way, and therefore evidence of its presence may appear at the latest
phase of the Tevatron collider and should be observed  in the next few years at
the LHC. However, the effective couplings of the lightest Higgs boson to gluons
and photons are modified by a light stop \cite{Djouadi:1998az}, and the
measurement of processes involving these couplings could provide evidence for
strongly interacting new particles with relatively low mass.

More challenging is the direct search for a scalar top quark. Acceptable values
of the relic density for heavy CP-odd Higgs bosons are naturally obtained in the
stop--neutralino co-annihilation region, in which the neutralino-stop mass
difference is of about a few tens of GeV~\cite{Csaba}.  In this case, the
three-body stop decay  channel into $W$ boson, bottom quark and the lightest
neutralino is kinematically closed, and the loop-induced two-body decay  into a
charm and a  neutralino tends to be the dominant stop decay
mode~\cite{Hikasa:1987db}. Searches for a stop decaying into a charm plus
missing energy at the Tevatron collider rely on a sufficiently energetic charm
quark, in order to trigger on these events. The charm transverse energy, in
turn,  is controlled by the mass difference of the stop and the neutralino. For
mass differences below 40~GeV, the Tevatron has no sensitivity in stop
searches~\cite{d0excl}\footnote{Very recently, however, it was proposed that the
reach of the Tevatron for small mass differences could be extended by not
requiring full reconstruction of the two charm
jets~\cite{Bhattacharyya:2008tw}.}. For larger mass differences, however, the
Tevatron reach extends up to values well above the LEP limit of 100 GeV,
allowing to explore stop masses up to 160~GeV~\cite{tevproj}.

Searches at the LHC in stop pair production are equally challenging. However,
at the LHC the stop may be produced in decays of heavy gluinos. Since the
gluino is a Majorana particle, it can decay into both a stop--antitop and
top--antistop final states. Pair production of gluinos may then lead to a pair
of equal-sign top quarks plus two equal-sign stops. It has been shown
that these events may be detected for gluino masses below
900~GeV~\cite{ak,Martin:2008aw}.

One would like to explore a method that would not depend so strongly
on the assumption of a relatively light gluino, which is not
required for either the generation of dark matter or the baryon
asymmetry. In this article we propose such a method: Stop particles
at the LHC may be produced in association with a hard photon or a hard jet.
In the co-annihilation region, there will be minimal hadronic
activity associated with the stop decays and therefore this would
effectively lead to events with a photon or a jet and missing energy.
Such a signature has been proposed, for instance, to explore extensions of the
Standard Model by large extra dimensions~\cite{led}. 
Recently, it has been pointed out that the jet plus missing energy signature 
can also be useful to search for relatively light gluinos at the
Tevatron~\cite{Alwall:2008ve}.
In this article we show that at the LHC these channels allow to test the
whole region of stop and neutralino masses consistent with
dark matter and electroweak baryogenesis (EWBG) that cannot be covered by
the Tevatron searches.

The article is organized as follows. In section \ref{sc:ewbg} we present an
overview of scenarios that accommodate a light stop. The following two sections
\ref{sc:get} and \ref{sc:jet} are devoted to phenomenological studies of the
photon plus missing energy and jet plus missing energy channels at the LHC. In
section \ref{sc:tev} we shortly comment about the relevance of these channels
for the Tevatron. Section \ref{sc:glu} compares our results to the reach for
stops in gluino decays at the LHC. Section \ref{sc:prop}
discusses some possibilities for identifying the stops from their
decay characteristics at the LHC. Finally, our conclusions are presented in
section \ref{sc:concl}.


\section{Scenarios with light scalar top quarks}
\label{sc:ewbg}

Although the particle content of minimal supersymmetric extensions of the 
Standard Model is well defined by symmetry principles, 
the exact spectrum depends strongly on the scale and
nature of the mechanism that leads to the soft supersymmetry breaking parameters.
Flavor physics puts strong constraints on the possible structure of supersymmetry
breaking terms. Strong flavor changing neutral current effects may be naturally
avoided if, at the messenger scale, at which supersymmetry breaking is transmitted
to the observable sector, the soft supersymmetry breaking terms are flavor 
independent. Under these assumptions all squarks and sleptons with 
the same quantum numbers 
under the SM gauge group will acquire equal masses at the messenger scale.

The observable supersymmetric particle masses may be obtained by renormalization
group evolution of the parameters at the messenger scale to the weak scale. 
There are different corrections appearing in this evolution~\cite{Ibanez:1984vq}. 
On one hand, strong
interaction effects, governed by the gluino mass, drive the squark masses to
larger values. There are also Yukawa induced effects that push the third generation
squark masses, in particular the stop masses, to smaller values.
These Yukawa effects are the same that drive
the effective Higgs mass parameter to negative values, inducing the breakdown of
the electroweak symmetry. Once the electroweak symmetry is broken,
there is also a Higgs induced mixing between the  right-handed and 
left-handed stops, that pushes the lightest stop to small values. In general,
in the scenarios with flavor independent scalar masses at the messenger scale, 
the lightest stop 
is the lightest squark, and the precise value
of its mass depends on the scale of supersymmetry breaking, the gluino mass 
and the size of the
stop mixing parameter. For relatively small values of the gluino 
mass and a large stop mixing
mass parameter, the lightest stop can have masses of the order or 
smaller than the top quark mass \cite{mssmft}. 
Due to the renormalization group
evolution such a light stop is dominantly the partner of the right-handed top
quark, which also concurs with electroweak precision constraints.

Light stops may also appear in alternative solutions to the flavor problem;
Since the strongest flavor physics constraints are associated with the first
two generations,  a simple solution is to let the first and second generation
squark masses  to be very large, while the stops remain light, in order to
prevent large radiative  corrections to the Higgs mass parameter. In these
``more minimal supersymmetry'' scenarios~\cite{CKN},  one considers that these
conditions are fulfilled at the weak scale, independently of the gluino mass or
the scale of supersymmetry breaking. Observe, however, that unless there are
additional interactions beyond the MSSM, one of the stops needs to be heavier in
order to prevent a Higgs mass below the experimental bound \cite{lephbound}. 

One can also consider a phenomenological scenario similar 
to split supersymmetry~\cite{split}, in
which all superparticles which remain light are there for a reason~\cite{Germano}. 
In this case,
a light stop would be necessary to induce  a strongly first-order phase transition,
necessary for electroweak baryogenesis, while light Higgsinos and gauginos would be necessary
to lead to a dark matter candidate, generate the necessary CP-violating currents and
for the model to be consistent with the unification of gauge couplings at high 
scales.

In all of the above scenarios, additional assumptions have to be made about the
sfermion flavor structure to predict the width and branching ratios of the
light stop. For this work, it is assumed that the two-body decay $\sta \to c +
\neu_1$ is enhanced by large logarithms from renormalization group evolution
\cite{Hikasa:1987db,Hiller:2008wp} so that this decay channel becomes dominant
and the stop decay length is too short to be observable. Alternative scenarios
where the partial width for $\sta \to c \, \neu_1$ is small lead to interesting
signals from the competing four-body decay $\sta \to b \, l^+ \nu_l \, \neu_1$
 and displaced stop decays
\cite{Hiller:2008wp} but will not be investigated further here.


\section{Stops in the \boldmath $\gamma+\Et$ channel}
\label{sc:get}

As explained in the previous section, we are interested in exploring
stop production in the co-annihilation region, where the mass
difference 
between the stop and the lightest supersymmetric particle,
which we assume to be the neutralino $\neu_1$,
\begin{equation} 
\Delta m = \mst{1} - \mneu{1},
\end{equation}
is of the order of a few tens
of GeV. In this region, the visible decay products of the stop tend to
be  soft, and therefore difficult to observe at hadron colliders.
However,
the lack of hadronic activity in the stop decay products may be
useful in searches for stops in other channels, for instance
by looking at the recoil of stops against a hard
photon or jet. In this section, we shall explore the possible
production of a stop in association with a hard photon.
Since the stop decays into relatively soft jets and missing
energy, the final state in the stop production channel  may be 
taken as  $\gamma+\Et$.
We have therefore performed the simulation of the signal 
\begin{equation}
p \, p \to
\sta \, \sta^* \, \gamma 
\end{equation}
using {\sc CompHEP 4.4} \cite{comphep}, interfaced with
{\sc Pythia 6.4} \cite{pythia} through the {\sc CPyth toolkit 2.0.6}
\cite{cpyth}.
Pythia has been run with power showers and including stop fragmentation
and hadronization
\cite{pythiastops} before decay $\sta \to c + \neu_1$.
The Pythia output has been fed into the fast detector simulation PGS~\cite{pgs},
in order to  simulate the most important detector effects.

\vspace{2ex}
The $\gamma+\Et$ signature has been considered previously for searches for large
extra dimensions at the LHC \cite{cmsgemiss,vacahinch}.
Therefore we can use the published results for the evaluation of the  
Standard Model background at the LHC. Our analysis is based on the SM
background estimates by the CMS collaboration in Ref.~\cite{cmsgemiss}.
The
main physics background channels come from the production of weak gauge
bosons, for instance, $\gamma Z$ with $Z\to\nu\bar{\nu}$, and $W \to e\nu$
where the electron is faking a photon.
These channels may be calibrated from observations in other well
measured production processes. For instance,
$\gamma Z$ with $Z\to\nu\bar{\nu}$ can be calibrated from
$\gamma Z$ with $Z\to l^+l^-$ with a total error of 3\% \cite{cmsgemiss}, 
using extrapolation from small $p_{\rm T,\gamma}$ to the signal region.

Since all colored particles, including gluinos and other squarks, 
are considered
to be heavy, with masses of about 1~TeV or larger, the potentially large SUSY 
contributions to the background associated with color particle production
are assumed to be negligible after applying
the analysis cuts. The main irreducible SUSY background stems from the
production of neutralinos with photons, which is, again, numerically
small and can be neglected compared to the SM backgrounds.

\vspace{2ex}
In order to obtain a reliable estimate of the expected stop signal significance
at the LHC, our
analysis has been performed using similar cuts as the ones
used in the CMS study, Ref.~\cite{cmsgemiss}: 
\begin{enumerate}
\item Require one hard photon with $p_{\rm T} > 400$ GeV and 
	pseudo-rapidity $|\eta| < 2.4$. 
\item Missing energy requirement: $\Et > 400$ GeV.
\item Veto against tracks with $p_{\rm T} > 40$ GeV.
\item Require back-to-back topology for photon and missing momentum:
$\Delta \phi(\ppslash_{\rm T},\vec{p}_\gamma) > 2.5$.
\item The photon has to be isolated. Ref.~\cite{cmsgemiss} uses a likelihood method
for photon isolation, but for simplicity we use the standard isolation 
criteria in
PGS~\cite{pgs}. The impact of these details on the signal is small.
\end{enumerate}
After applying these cuts, the remaining SM background is relatively
small, about 2.5~fb, corresponding to 
250 events for 100~fb$^{-1}$ \cite{cmsgemiss}.

\vspace{2ex}
The above cuts, however, also affect the signal rate, particularly for
increasing values of $\Delta m$ due to the strong $\Et$ requirement 
established above. One could in principle optimize the cuts as a function
of $\Delta m$, but this would require a reevaluation of background, which is 
beyond the scope of this paper.
Since no cut optimization has been performed, the
procedure we are applying serves therefore as a conservative estimate
of the reach in the stop--neutralino mass parameter space.

\vspace{2ex}
Although the SM backgrounds are known including next-to-leading
(NLO) QCD corrections,
no such corrections are available for the $\sta\sta^*\gamma$ process.
In our analysis, we shall assume  a
K-factor of 1.4, an estimate that comes from the calculation of the process 
$pp \to \sta\sta^*$ \cite{nlolhc}.
However, this value should be taken with caution since
the veto of any hard real radiation 
jets due to our selection cuts could reduce the effective size of the radiative
corrections, an effect that is partially but not precisely reproduced by the
parton shower in {\sc Pythia}\footnote{Due to the missing NLO corrections,
our evaluation of the $\gamma+\Et$ channel has sizable theoretical
uncertainties. However, the jet$+\Et$ channel, discussed in the next section, gives much
better and more robust results.}.

\begin{figure}[tb]
\centering
\epsfig{figure=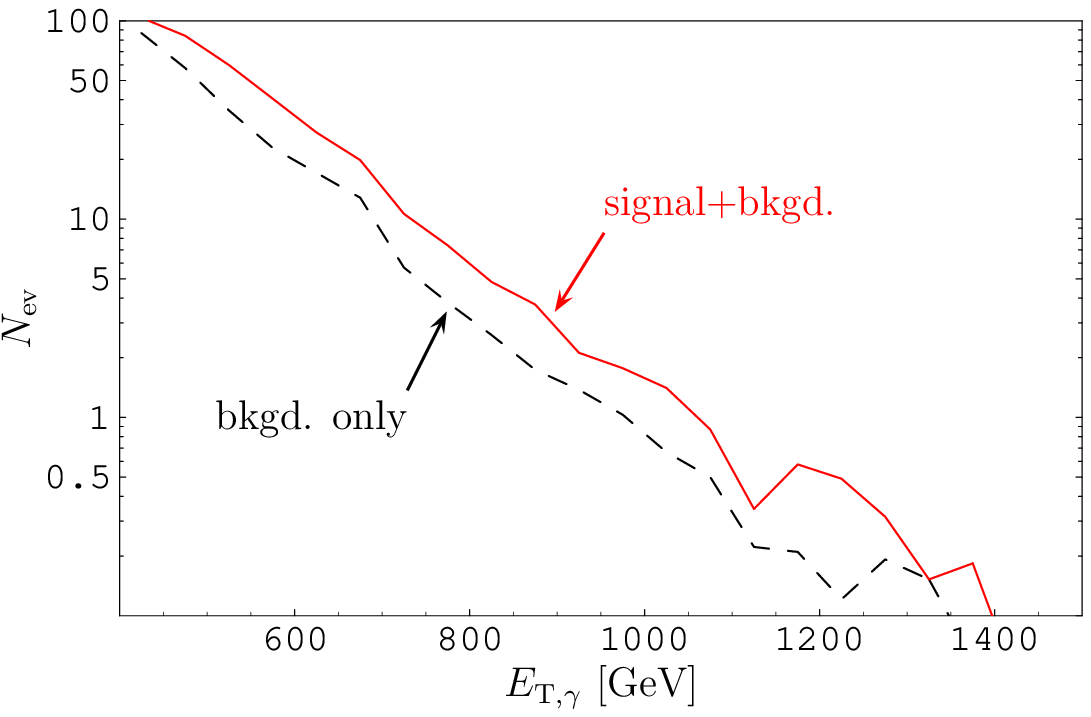, width=10cm}
\mycaption{Distribution of the photon transverse energy, $E_{\rm T,\gamma}$,
for the SM background (from Ref.~\cite{cmsgemiss}) and the stop signal with
$\mst{1}$=130~GeV and $\mneu{1}$=110~GeV. The fluctuations at the right end of
the plot are due to Monte Carlo errors.}
\label{fg:etg}
\end{figure}

\vspace{2ex}
After application of the cuts and the K-factor rescaling the signal event number
for 100~fb$^{-1}$ is shown in Tab.~\ref{tab:photon_sig} for various values of
the stop and neutralino masses.
\begin{table}[tb]
\centering
\begin{tabular}{|r|rrrrrrr|}
\hline
 \multicolumn{2}{|r}{$\mst{1}/\mbox{\rm GeV} =110$} & 130 & 150 & 170 & 190 & 210 & 230 \\
\hline
$\Delta m/\mbox{\rm GeV}=10$ & $\quad$189 & 173 & 157 & 138 & 114 & 103 & 85 \\
20 & 111 & 119 & 110 & 99 & 86 & 81 & 71 \\
30 & 72 & 74 & 77 & 80 & 64 & 60 & 57 \\
40 & 56 & 54 & 55 & 53 & 48 & 45 & 45 \\
50 & 43 & 43 & 40 & 40 & 37 & 35 & 34 \\
\hline
\end{tabular}
\mycaption{Number of signal events in the $\gamma+\Et$ channel
for 100~fb$^{-1}$ and for various combinations of
$\mst{1}$ and $\Delta m = \mst{1}-\mneu{1}$.
The numbers in the table have an intrinsic statistical uncertainty of a few
events from the Monte Carlo error.}
\label{tab:photon_sig}
\end{table}
Figure~\ref{fg:etg} shows the distribution of the photon transverse energy
for the SM background
and the stop signal for $\mst{1}=130$~GeV and $\mneu{1}=110$~GeV ({\it i.e.}
$\Delta m = 20$~GeV).
As can be seen from the figure, the shape of the $E_{\rm T,\gamma}$ distribution 
is very similar for the background and stop signal, so that the only
discriminating feature is the overall event count.

\begin{figure}[tb]
\centering
\epsfig{figure=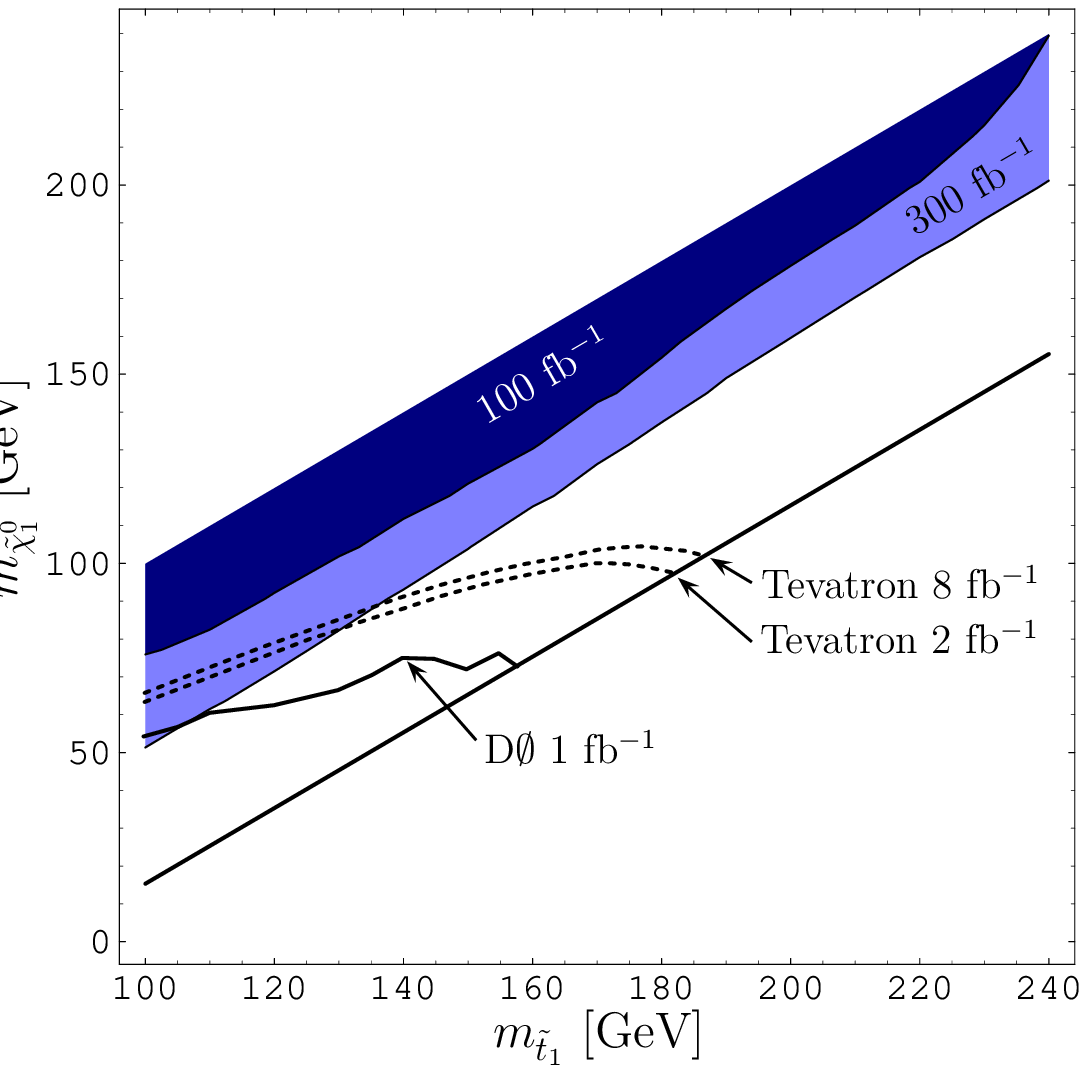, width=8cm} \ 
\epsfig{figure=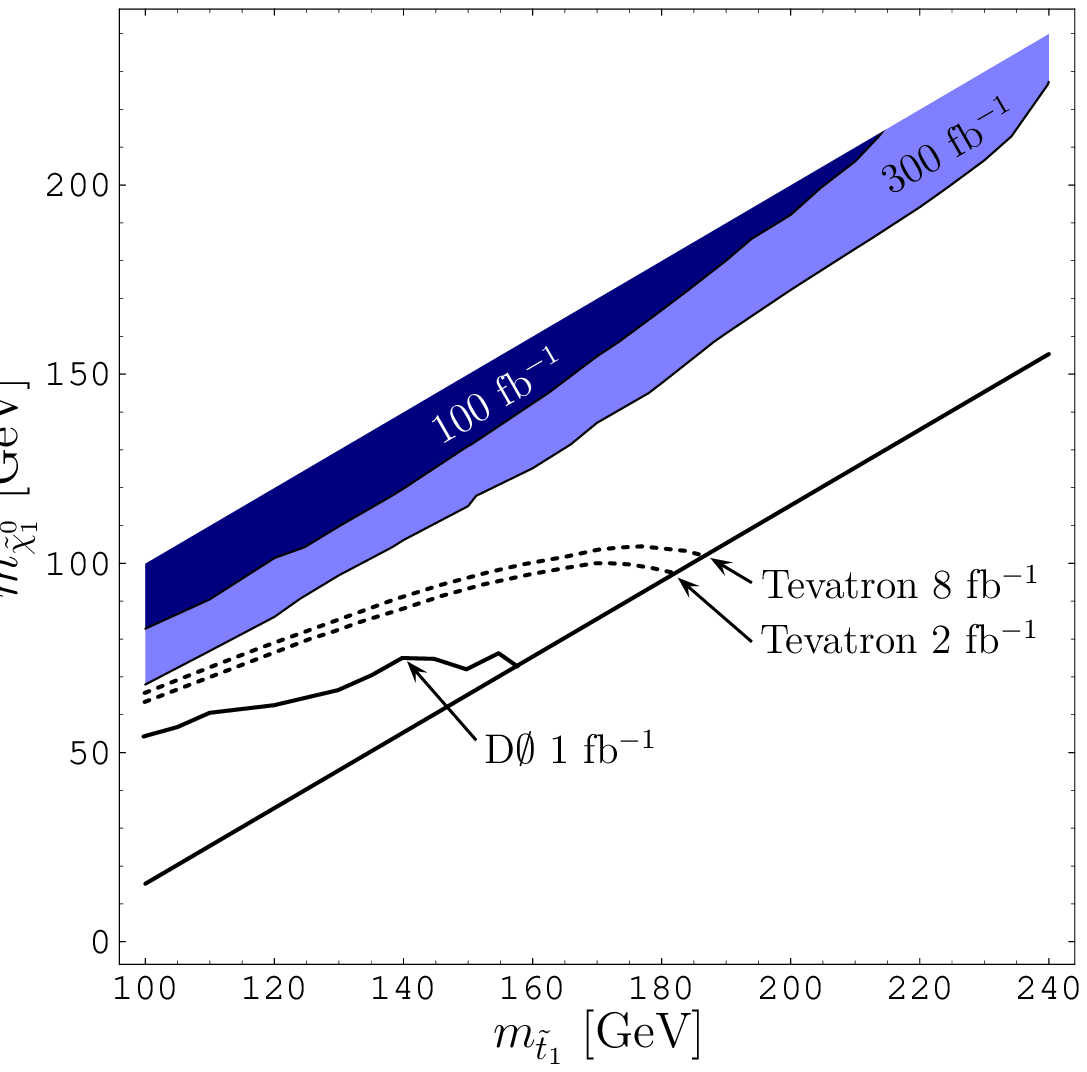, width=8cm}
\mycaption{Projected LHC 5$\sigma$ discovery reach in the $\gamma+\Et$ channel without (left) and with (right) systematic
errors. For comparison the current and future Tevatron 95\% C.$\,$L.\ exclusion bounds 
for light stops are also shown.}
\label{fg:gemiss}
\end{figure}

Using this information,
the projected reach for stop searches by means of the procedure explained
above is represented in
Fig.~\ref{fg:gemiss} without (left) and with (right) systematic
errors. In Fig.~\ref{fg:gemiss} 
we have considered the following systematic errors:
\begin{itemize}
\item Assuming a
2\% error on the measaurement of the photon $p_{\rm T}$ leads to a
3\% uncertainty of the SM background.
The estimate has been obtained by simulating events for the dominant 
background $\gamma Z$ with $Z\to\nu\bar{\nu}$ and it agrees well with
the evaluation in Ref.~\cite{cmsgemiss}.
\item Assuming a 5\% error on the measurement of $\Et$,
we obtain an effect of 5\% on the background, which is slightly larger than
the number of 4\% quoted in  Ref.~\cite{cmsgemiss}.
\item 
A precise determination of parton distribution functions (PDFs)
is crucial for the
measurement of the $\gamma+\Et$ cross section. We expect that the PDFs can be
constrained by measurements of the reference process $\gamma Z$ with $Z\to
l^+l^-$. Therefore the systematic error due to PDFs
should be similar to the statistical error of that process, {\it i.e.} about
3\%.
\end{itemize}
Adding the individual contributions in quadrature,
the total systematic error is expected to be about~6.5\%.

Figure~\ref{fg:gemiss} shows the 5$\sigma$ discovery reach with the statistical
significance estimated by $S/\sqrt{B}$.
Also shown are current~\cite{d0excl} and projected~\cite{tevproj} 95\%
confidence level (C.$\,$L.) exclusion limits
for light stop searches in the jets plus missing energy channel
at the Tevatron collider.
Observe that since Tevatron searches are more efficient for large values
of $\Delta m$, while the LHC searches explored in this article become
more efficient for small values of $\Delta m$, searches for stops at
both colliders complement each other. High luminosities
at both hadron colliders allow to probe a large fraction of the region of 
stop--neutralino masses consistent with electroweak baryogenesis, 
$m_{\tilde{t}_1} < 130$~GeV. This is particularly so in the
co-annihilation region, but can be applied to larger values of 
$\Delta m$ if the systematic errors discussed above were reduced,
getting results close to the left panel of Fig.~\ref{fg:gemiss}.


\section{Stops in the \boldmath jet$+\Et$ channel}
\label{sc:jet}

\noindent
The search for photons and $\Et$ explored in the last
section could be a promising way of looking for stops in
the co-annihilation region. However, the reach is limited due to the
relatively small signal cross section.
An alternative method, with similar
properties as the searches discussed above is to look for
the recoil of stops against hard jets.
An advantage of this search channel compared to the 
$\gamma+\Et$ is the much larger rate 
induced by the strong interaction production process. A clear
disadvantage, however, is related to larger measurement uncertainties, 
and increased backgrounds and systematic errors. In particular, to control
the potentially large QCD background, large missing transverse energy 
$\Et > 1$~TeV must be demanded.

\vspace{2ex}
Due to the small hadronic activity associated with the decaying stop,
and analogously to the photon case, the main signature of this process
is
\begin{equation}
p p \to {\rm jet} +\Et.
\end{equation}
As in the photon case, we have performed a Monte Carlo simulation of this 
discovery channel.
The signal $\sta\sta^*+j$ 
has been generated
using {\sc CompHEP 4.4} \cite{comphep}, interfaced with
{\sc Pythia 6.4} \cite{pythia}. 
No matching procedure has been applied for 
jet radiation from matrix elements and parton showers, 
but errors due to that should be small since the typical $p_{\rm T}$ 
for the hardest jet is very large to balance the large $\Et$.

\vspace{2ex}
As in the photon case, we have extracted the SM backgrounds from
previous experimental studies at the LHC. The jet+$\Et$ channel
has been investigated in searching for large extra dimensions in 
Ref.~\cite{vacahinch}, which also contains a detailed analysis of
SM backgrounds.
The main physics background channels are also similar to the photon
case : $jZ$ with $Z\to\nu\bar{\nu}$,  and $jW$ with $W \to \tau\nu$.
Here $j$ stands for a hard jet.
The process
$jZ$ with $Z\to\nu\bar{\nu}$ can be calibrated from
$jZ$ with $Z\to l^+l^-$ \cite{vacahinch}, and for similar reasons
as in the photon case, the SUSY background has been assumed to be small.

In order to proceed with this analysis, we have used the same cuts as 
in Ref.~\cite{vacahinch}: 
\begin{enumerate}
\item Require one hard jet with $p_{\rm T} > 100$ GeV and $|\eta| < 3.2$ for the
trigger. 
\item Large missing energy $\Et > 1000$ GeV.
\item Veto against electrons with $p_{\rm T} > 5$ GeV and muons with
$p_{\rm T} > 6$ GeV in the visible region ($|\eta| < 2.5$).
\item Require the second-hardest jet to go in the opposite hemisphere as 
the missing momentum ({\it i.e.}
the first and second jet should go in roughly the same direction):
$\Delta \phi(p_{\rm T,j_2},\vec{p}_\gamma) > 0.5$.
This cut reduces background from $W\to \tau\nu$ where the tau decay products are
emitted mostly in the opposite direction as the hard initial-state jet.
\end{enumerate}
Application of these cuts leads to a
SM Background of about 7~fb, 
corresponding to 
700 events for 100~fb$^{-1}$ \cite{vacahinch}.

\vspace{2ex}
The 
NLO corrections to $\sta\sta^* + j$ are not available
in the literature. However, experience from $t\bar{t}j$~\cite{Dittmaier:2007wz}
suggests that the K-factor should be  close to one. 
Therefore, contrary to what was done in the photon
case, we shall not include a K-factor for the signal.

\begin{figure}[tb]
\centering
\epsfig{figure=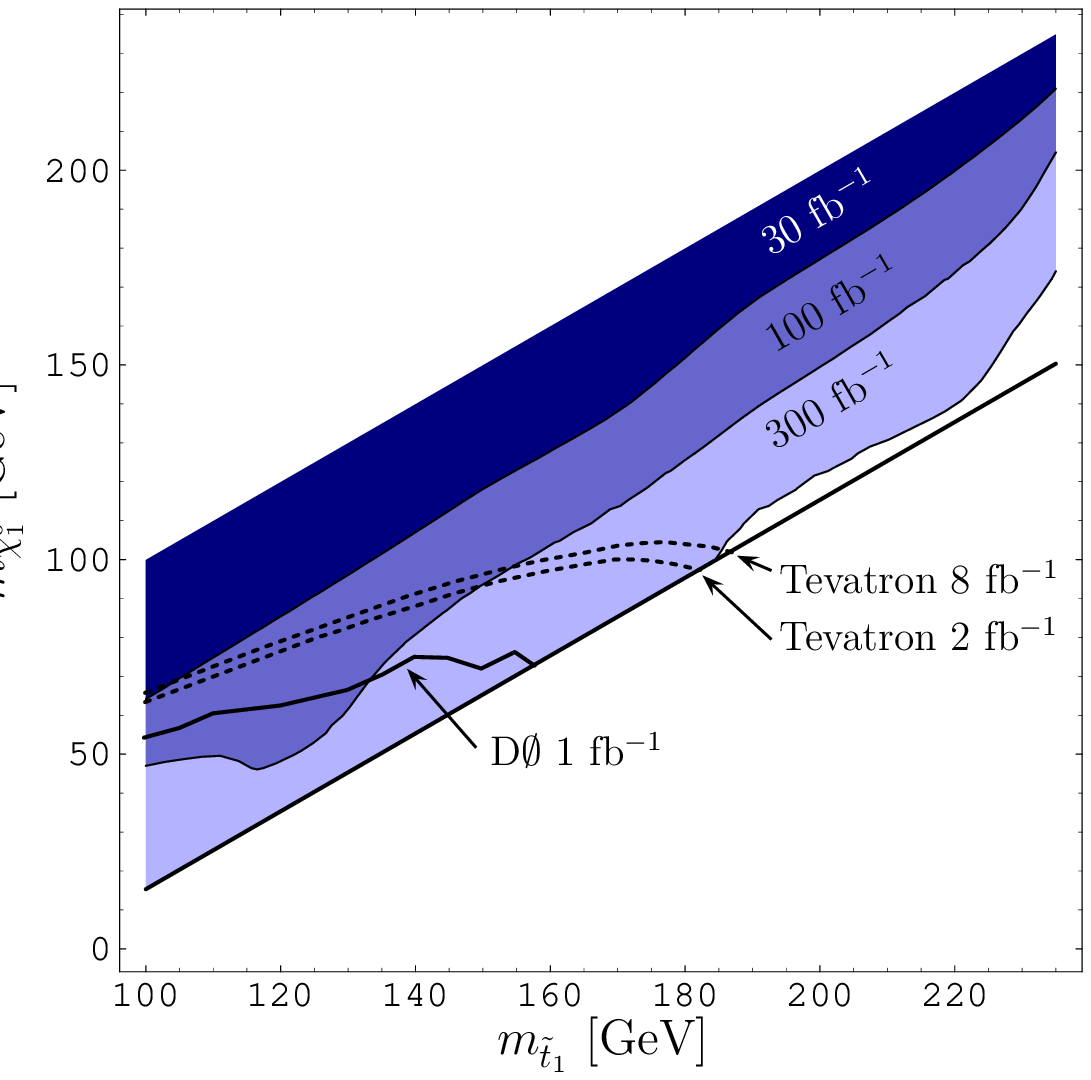, width=8cm}
\mycaption{Projected LHC 5$\sigma$ discovery reach in the jet$+\Et$ channel. For
comparison the current and future Tevatron 95\% C.$\,$L.\ exclusion bounds 
for light stops are also shown.}
\label{fg:jemiss}
\end{figure}

\begin{table}[tb]
\centering
\begin{tabular}{|r|rrrrrrr|}
\hline
 \multicolumn{2}{|r}{$\mst{1}/\mbox{\rm GeV} =110$} & 130 & 150 & 170 & 190 & 210 & 230 \\
\hline
$\Delta m/\mbox{\rm GeV}=10$ & $\quad$%
     1920 & 1716 & 1585 & 1360 & 1056 & 1015 & 845 \\
20 & 1170 & 1085 & 948 & 877 & 717 & 676 & 570 \\
30 & 762 & 746 & 676 & 679 & 548 & 551 & 433 \\
40 & 559 & 516 & 514 & 507 & 442 & 444 & 348 \\
50 & 437 & 449 & 422 & 428 & 364 & 343 & 279 \\
\hline
\end{tabular}
\mycaption{Number of signal events in the jet$+\Et$ channel
for 100~fb$^{-1}$ and for various combinations of
$\mst{1}$ and $\Delta m = \mst{1}-\mneu{1}$.
The event numbers in the table have an intrinsic statistical uncertainty of a few
tens from the Monte Carlo error.}
\label{tab:jet_sig}
\end{table}
\vspace{2ex}
Using the above defined cuts, the expected number of signal events is listed in 
Tab.~\ref{tab:jet_sig} for various stop and neutralino mass values.
Fig.~\ref{fg:jemiss} 
shows the projected 5$\sigma$ discovery reach with the statistical
significance estimated by $S/\sqrt{B}$ and including systematic errors.
In order to estimate the systematic errors, we have explored the following
two strategies, (a) and (b): 
\begin{enumerate}
\item[(a)]
The first strategy determines the dominant SM backgrounds directly from 
data~\cite{vacahinch}. In particular,
the $jZ$ background
with $Z\to\nu\bar{\nu}$, which contributes about 75\% of the SM background
after cuts, can be inferred from $jZ$ with $Z\to l^+l^-$,
$l=e,\mu$. The $Z\to l^+l^-$ calibration channel is about seven times smaller
than the $Z\to\nu\bar{\nu}$ background in the signal region 
($p_{{\rm T},ll} > 1$~TeV), 
thus leading to the error estimate $\delta_{\rm sys} B = \sqrt{7B}$.
\item[(b)]
Alternatively, similar to the previous section,
individual systematic error sources can be identified:
\begin{itemize}
\item A 5\% error on $\Et$ induces a 36\% uncertainty on the background,
	as determined by simulating $j Z$ with $Z\to\nu\bar{\nu}$.
\item The PDFs can be extracted from reference SM processes, {\it e.g.}
 $jZ$ with $Z\to l^+l^-$. Thus the uncertainty is mainly limited 
 by the statistical error for the standard candle process.
 For the region of high transverse momenta
 ($p_{\rm T} > 500$~GeV), which is relevant for the present analysis,
 this leads to relatively small error of 3\%.
\item Systematic uncertainties associated with the
 lepton veto are negligible, since this cut plays a role mainly for the 
$jW$ background with $W \to e\nu$ or $W \to \mu\nu$, which contributes only
about 5\% to the total SM background.
\end{itemize}
In summary, this strategy yields a
total estimated systematic error of about 36\%, strongly dominated by the
uncertainty of the missing $\Et$ measurement.
\end{enumerate}
It is evident that the data-driven method (a) for determining the systematic
error of the SM backgrounds leads to better results.
This is different from the photon case in section~\ref{sc:get}, in which
method (b) proves to be convenient. The improvement in
the results associated with method (a) in the jet
case is due to the larger statistics, while on the other hand
a much larger background uncertainty is induced for method (b)
by the error in the missing energy determination.

The results presented in  Fig.~\ref{fg:jemiss} make use of method (a).
Searches in the jet plus $\Et$ channel turn out to be more promising than in
the photon plus $\Et$ channel. They allow to test the co-annihilation
region up to relatively large values of the stop mass, of about 200~GeV
or larger. Moreover, when complemented with Tevatron search analyses, they
allow to fully explore the region of stop masses consistent
with electroweak baryogenesis with only 100 fb$^{-1}$  of integrated
luminosity.


\section{Stops in \boldmath $\gamma+\Et$ and jet$+\Et$ at the Tevatron}
\label{sc:tev}

In principle, the $\gamma+\Et$ and jet$+\Et$ channels could be used
already at the Tevatron for searching for stops with small stop-neutralino
mass difference, a region of parameter space which is difficult to access
with traditional search strategies.

Using {\sc CompHEP 4.4} \cite{comphep}, we have computed the stop signal cross
section for the Tevatron in these channels, and have compared them to the
background evaluations by the CDF~\cite{cdfed} and D$\emptyset$~\cite{d0ed}
collaborations.

For the  $\sta\sta^* + \gamma$ channel with a stop mass of $\mst{1} = 100$~GeV,
the Tevatron cross section for $p_{\rm T,\gamma} > 90$~GeV and $|\eta_\gamma| <
1$ is about 3.2 fb, which is of the same order as the systematic error in the
background analysis of CDF ($\delta_{\rm sys} = 1.5$~fb)
and D$\SLASH{0}{1.2}$ ($\delta_{\rm sys} = 1.5$~fb).
For larger values of $\mst{1}$ the signal cross section is even smaller.

In the  $\sta\sta^* + j$ channel with $\mst{1} = 100$~GeV and the minimal cut
$p_{\rm T,j_1} > 150$~GeV the Tevatron cross section is about 50~fb, which is
smaller than the estimated systematic error on the SM background of 56~fb
\cite{cdfed}.

Our conclusion is that the Tevatron will not be able to discover stops via the
$\gamma+\Et$ or jet$+\Et$ channels. However, searches in the $\gamma+\Et$
signature could
exclude light stops with $\mst{1} \sim 100$~GeV at the 95\% confidence level. A
final statement about exclusion limits would require a more detailed experimental
analysis.


\section{Stops in gluino decays}
\label{sc:glu}

As has been proposed in Refs.~\cite{ak,Martin:2008aw}, 
if gluinos are light enough, stops can be discovered in their decays.
Due to the Majorana nature of gluinos, they may decay in two
CP-related channels, 
\begin{equation}
\tilde{g}
\to \sta \bar{t}, \; \sta^* t.
\end{equation}
One can therefore make use of this property to look for same-sign
top quark signatures (using leptonic $W$ decays) plus missing energy
in gluino pair production processes. Same-sign top quark  channels 
have much smaller backgrounds than the opposite-sign top quark
processes, and allow an efficient search for light stops for relatively
light gluinos. 

\vspace{2ex}
For the sake of comparison to our results in the previous sections, 
in this section, we re-evaluate the LHC stop discovery
reach in this process, using the same cuts as in Ref.~\cite{ak}:
\begin{itemize}
\item Two same-sign leptons with $p_{\rm T} > 20$ GeV.
\item At least {\it (a)} two or {\it (b)} four jets with $p_{\rm T} > 50$ GeV.
The two-jet selection (a) preserves more of the signal for small $\Delta m$, 
while the
four-jet selection (b) gives a better signal-to-background ratio for $\Delta m
\gesim 10$ GeV.
For a given MSSM scenario, we always choose the selection method (a) or (b) 
which gives a better signal significance.
\item At least 2 b-tagged jets with $p_{\rm T} > 50$ GeV. It has been assumed
that the b-tagging efficiency is 43\% per bottom jet, while the mis-tagging
rates are 10\% for charm jets and 2.5\% for light-flavor jets.
\item $\Et > 100$ GeV.
\item Two combinations of lepton and b-jet momenta have to give 
$m_{bl} < 160$ GeV, in order to reduce non-top background.
\end{itemize}
Using {\sc Pythia 6.4} \cite{pythia} interfaced with PGS \cite{pgs}, we were
able to reproduce the signal numbers in Ref.~\cite{ak} within Monte Carlo errors.

\begin{figure}[tb]
\centering
\epsfig{figure=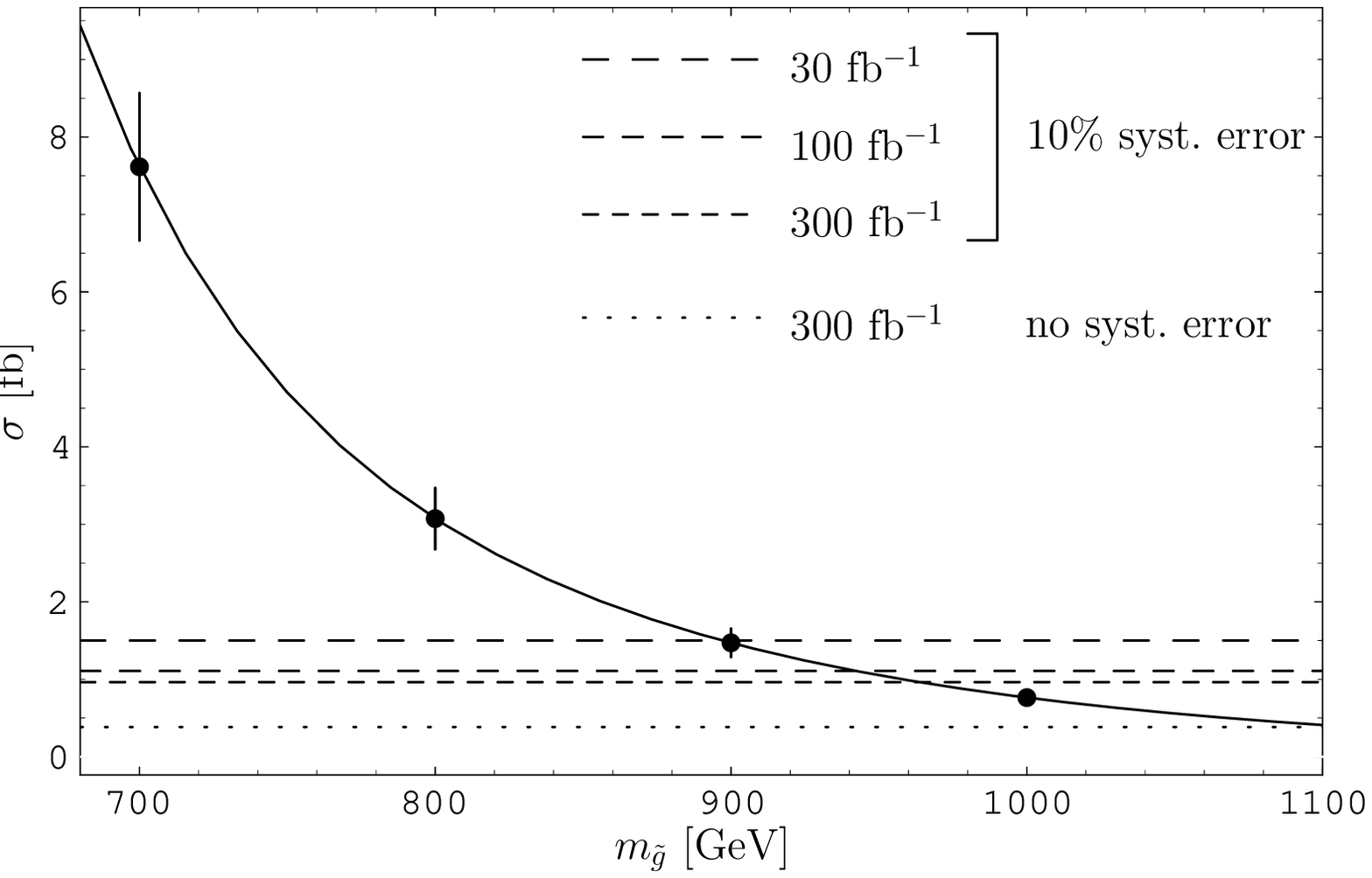, width=10cm}
\mycaption{Projected LHC reach in the $\tilde{g} \tilde{g} \to 
t t \sta^* \sta^* \; (\bar{t} \bar{t} \sta \sta)$ channel.
The errors bars indicate Monte Carlo errors.}
\label{fg:gluino}
\end{figure}

Scanning over a wide range of sparticle masses, we found that 
the expected discovery reach of the LHC in this channel depends only mildly on
stop and neutralino masses, but strongly on the gluino mass. 
In Figure~\ref{fg:gluino}, we present the results of our analysis.
These results suggest that, as already stated in Ref.~\cite{ak},
for 30~fb$^{-1}$,
the stop reach in this channel extends to about $m_{\tilde{g}} = 900$ GeV.
Higher luminosities at the LHC allow to slightly extend the region of 
gluino masses, but, after considering systematic errors, still gluino
masses $m_{\tilde{g}} < 1$~TeV would be required for an efficient search
for stops in this channel. 
Here we have assumed a systematic error of 10\% on the remaning SM background
after cuts, which is dominated by $t\bar{t}$. The major systematic uncertainty
for this background stems from the measurement of $\Et$; a 5\% error on
$\Et$ induces an uncertainty of 10\% on the $t\bar{t}$ rate.


\section{Stop identification at the LHC}
\label{sc:prop}

\noindent
In the previous sections we have analyzed the possible searches for stops in
associated production with hard photons or jets at the LHC. If an excess in
these channels were observed, it would be very important to be
able to determine that indeed stops, and not other particles, 
are the source of the  missing energy events. In order to do
this, one would have to detect the relatively soft charm jets coming from
the stop decay $\tilde{t}_1 \to c \, \neu_1$.%
\footnote{At an $e^+e^-$ collider, a detailed analysis of stop decays and other
properties is possible with high precision \cite{stop}, but here we want to focus on
measurements at the LHC alone.}

In the following we shall attempt to identify the charm-jets by means of  jet
mass and track multiplicity. We focus on the $\gamma+\Et$ study as an example,
and look at the soft jets that survive the track veto and other selection
cuts. For each jet its mass is calculated as the invariant mass of the momentum
vectors associated with the calorimeter hits inside the jet cluster.

\begin{figure}[tb]
\centering
\epsfig{figure=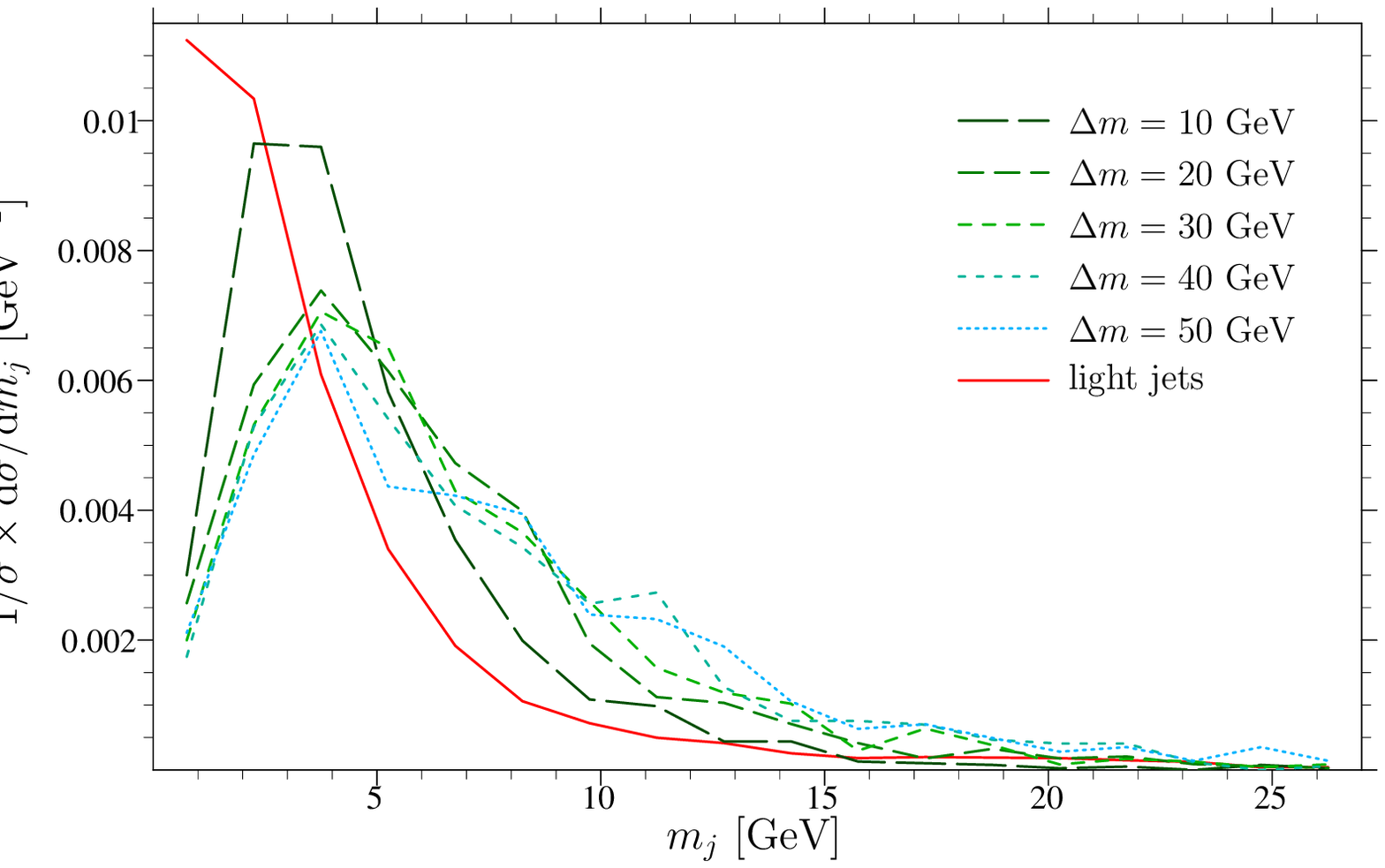, width=10cm}
\mycaption{Jet mass distribution for charm jets from
stop decays for different $\Delta m$, compared to light flavor jets from ISR.}
\label{fg:jmass}
\end{figure}

Fig.~\ref{fg:jmass} shows the 
jet mass distribution for charm jets from
stop decays for different values of 
$\Delta m$, and for light-flavor jets from initial state radiation (these
light ISR jets come both from signal and background).
As evident from the figure, the jet mass distribution is clearly different
for light-flavor  jets and charm jets.
Cutting $m_j > 4.5$ GeV keeps about 60\% of the charm 
jets (for $\Delta m \gesim 20$
GeV) but only 25\% of light flavor jets. 

The distinction between light-flavor and charm jets from the jet mass
becomes difficult for very
small mass differences $\Delta m \sim 10$ GeV.
However, the charm tagging performance can be improved by including other variables
in addition to the jet mass. The implementation of a state-of-the-art 
charm tagging algorithm is beyond the scope of this work, but we have designed a
simple two-variable tagging method using the jet mass and track multiplicity
within the jet. Track multiplicity as a discriminatory variable is particularly
useful for small $\Delta m$ since in this case the charm jet contains
fewer charged tracks than a typical light-flavor jet. This can be explained by
the limited phase space available for QCD radiation from a soft charm quark.
The results for the tagging efficiency are shown in
Tab.~\ref{tab:ctag}.
\begin{table}[tb]
\centering
\begin{tabular}{|l|ccccc|c|}
\hline
 & \multicolumn{5}{|c|}{Charm jets} & Light-flavor \\
\cline{1-6}
$\Delta m$ [GeV] & 10 & 20 & 30 & 40 & 50 & \raisebox{.2ex}{jets} \\
\hline
Efficiency & 50\% & 60\% & 63\% & 65\% & 66\% & 25\% \\
\hline
\end{tabular}
\mycaption{Charm tagging efficiency and light-jet mistagging rate for a simple
tagging algorithm based on jet mass and track multiplicity.}
\label{tab:ctag}
\end{table}

%
%
As mentioned above, charm tagging can be used to identify the flavor of the stop decay products.
As an example,  we have chosen the following
sample parameter point: $\mst{1} = 130$ GeV, $\Delta m = 20$ GeV. 
The signal can be
detected with $>5\sigma$ with 100 fb$^{-1}$ for this point, yielding 119 signal
and 251 background events. If only light-flavor jets were present in the entire
sample, the requirement of at least one charm-tagged jet with $p_{\rm T} > 20$ 
GeV would reduce the event count to 23\%.
In reality, due to the charm jets coming from stop decays, 31.5\% survive. 
This
means that the presence of heavy flavor jets in the signal can be inferred
experimentally with 2.9$\sigma$. With 300 fb$^{-1}$, this improves to
5$\sigma$. 

\begin{figure}[tb]
\centering
\epsfig{figure=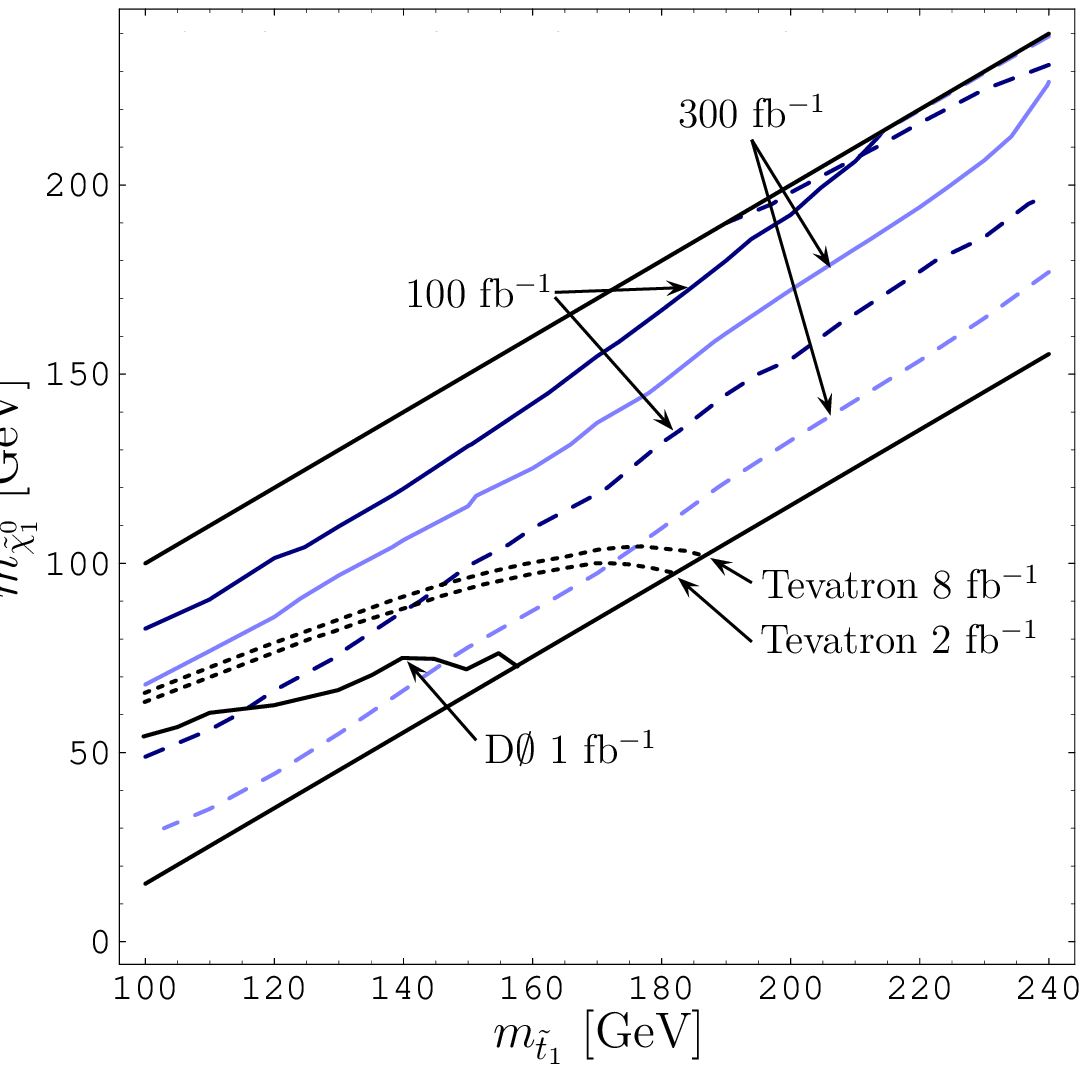, width=8cm}
\mycaption{Improvement of projected LHC reach in the $\gamma+\Et$ channel
from charm tagging,
for 100~fb$^{-1}$ (dark lines) and 300~fb$^{-1}$ (light lines).
The solid lines correspond to the right side of Fig.~\ref{fg:gemiss}
which has no charm tagging, while the dashed lines indicate the extended reach
due to charm tagging as described in the text.}
\label{fg:ctag}
\end{figure}

\vspace{2ex}
Apart from allowing to determine the flavor of the stop decay product, charm
tagging can also improve the stop discovery reach compared to the analysis in
section~\ref{sc:get}, where the decay products of the stops did not play any
role in the signal selection.
Here, in addition to the cuts in
section~\ref{sc:get}, we demand at least one charm-tagged jet with $p_{\rm T} >
20$ GeV, using the tagging efficiencies in Tab.~\ref{tab:ctag}. With this additional
cut, we find that the  region bounded by the dashed lines in Fig.~\ref{fg:ctag}
becomes accessible in the  $\gamma+\Et$ channel. As evident from the figure, the
discovery region is greatly extended compared to the results from
section~\ref{sc:get}.

Nevertheless, the treatment of charm tagging in this work is rather simple and
rudimentary, since we are not using a full detector simulation. A detailed
detector simulation is necessary to evaluate information about displaced
vertices, impact parameters and tagged mesons. With a combination of all
these variables more sophisticated charm tagging algorithms can be designed
which could yield even better results than our analysis.


Once the charm jets are identified one could in principle use kinematical
measurements to obtain information about the stop and neutralino masses, see
for example Ref.~\cite{mt2}. However, for the $\gamma+\Et$ and jet$+\Et$
channels such an analysis is very difficult due to the relatively small
signal-to-background ratio and distortions stemming from the rejection of extra
jets in the selection cuts. A conclusive statement about the measurability of
the masses would require a detailed study with proper simulation of the
backgrounds, which is beyond the scope of this paper.

\section{Conclusions}
\label{sc:concl}

In this article, we have investigated the possible search for light stops in
their production in association with hard photons or jets. Light stops may
naturally appear in many SUSY scenarios and are required to realize the
mechanism of electroweak baryogenesis in the MSSM. These search channels are
particularly useful in the case of small mass differences $\Delta m$ between the
stop and the neutralino, for which the jets coming from the decay of the stops
are relatively soft. Such small mass differences enable the stop-neutralino
co-annihilation process, yielding a dark matter density consistent with
experimental observations.

Searches
for pair of stops in two jets plus $\Et$ channels at the Tevatron and the
LHC require sufficiently hard jets in order to trigger on these events.
For small mass differences between the stop and the neutralino these
searches become therefore very challenging. The lack of hadronic
activity in the stop decay products may be used as an advantage in
processes in which the stops are produced in association with 
photons or jets, since effectively they can lead to final states
with one hard photon or jet and missing energy. In this article,
we have analyzed such processes. We showed that the photon plus
$\Et$ channel may be used to explore most of the light stop 
co-annihilation region, but it becomes rapidly inefficient for 
relatively large mass
differences $\Delta m$, particularly due to the existence of
large systematic errors. Searches in the jet plus $\Et$ channel,
instead, profit from larger production rates and lead to an
extended coverage of the light stop--neutralino mass plane. These
searches may be complemented with the Tevatron searches, which
become more efficient for larger values of $\Delta m$. Together,
for an LHC luminosity of about 100~fb$^{-1}$,
they cover all of the parameter space consistent with electroweak
baryogenesis, independent of the value of the
gluino mass.
Since we have used existing experimental simulations for the background
evaluation the selection cuts are not optimized for our signal process.
An optimization of the cuts thus
could further improve our results.

We have also reanalyzed the possibility of looking for stops in events
with a pair of like-sign top quarks proceeding from Majorana gluino decays,
and have confirmed the results of previous analyses which showed that
for an LHC luminosity of 30 fb$^{-1}$,
these searches become efficient for gluino masses below about 900~GeV.
Relatively light gluinos, with masses below about 1~TeV, are required
even for an LHC luminosity of about 300 fb$^{-1}$.

In the searches described above the stop properties are not well identified in
these processes. Therefore, we have also analyzed the possibility of identifying
stops, once an evidence of their presence in some of these processes has been
observed, by identifying the associated charm jets and showed that this is
possible for mass difference $\Delta m $ above about 10 GeV. A determination of
the stop and neutralino masses from the jet/photon  plus $\Et$ channels seems
very challenging. In the future, a more detailed study with a realistic detector
simulation should be performed to confirm our results.


\section*{Acknowledgments}

We would like to thank S.~Kraml, A.~Raklev, M.~Schmitt, T.~Sj\"ostrand and
J.~Wacker for useful discussions and comments. Work at ANL is supported in part
by  the US DOE, Division of HEP, Contract DE-AC-02-06CH11357.  Fermilab is 
operated by Universities Research Association Inc. under contract no. 
DE-AC-02-76CH02000 with the DOE.  The authors are thankful to the Aspen Center
for Physics, where part of this work has been performed.  M.~C.\ and C.~W.\ are
grateful to the KITP, Santa Barbara, and the KITPC, Beijing, for hospitality
during stages of this work.

\end{document}